\documentclass[USenglish]{article}	

\usepackage[utf8]{inputenc}				
\usepackage[big,online]{dgruyter}	
\usepackage{lmodern} 
\usepackage{microtype}
\usepackage[numbers,square,sort&compress]{natbib}
\usepackage{appendix}
\usepackage{xcolor}

\theoremstyle{dgthm}

\theoremstyle{dgdef}

\begin{document}

\articletype{Research Article}
\startpage{1}

\title{Where to serve and return in Badminton Men's Double?}
\runningtitle{Where to serve and return in Badminton Men's Double?}

\author[1]{Xuelin Zhu}
\author[2]{Yu Sun} 
\author[3]{Yumin Zeng} 
\author*[4]{Cong Xu}
\affil[1]{\protect\raggedright 
University of Michigan, Department of Statistics, Ann Arbor MI, U.S., e-mail: xuelin@umich.edu}
\affil[2]{\protect\raggedright 
Southern University of Science and Technology, Sports Center, Shenzhen, China, e-mail: suny3@sustech.edu.cn}
\affil[3]{\protect\raggedright 
Southern University of Science and Technology, Sports Center, Shenzhen, China, e-mail: zengym@sustech.edu.cn}
\affil[4]{\protect\raggedright 
Southern University of Science and Technology, Department of Statistics and Data Science, Shenzhen, China, e-mail: xuc6@sustech.edu.cn}

	
\abstract{This study aims to analyze the service and return landing areas in badminton men's double, based on data extracted from 20 badminton matches. We find that most services land near the center-line, while returns tend to land in the crossing areas of the serving team's court. Using generalized logit models, we are able to predict the return landing area based on features of the service and return round. We find that the direction of the service and the footwork and grip of the receiver could indicate his intended return landing area. Additionally, we discover that servers tend to intercept in specific areas based on their serving position. Our results offer valuable insights into the strategic decisions made by players in the service and return of a badminton rally.}


\maketitle

\section{Introduction} 
Badminton is a popular racquet sport played all over the world, particularly in Asia. The goal of the game is to score points by striking the shuttlecock (commonly referred to as the shuttle) so that it lands within the opponent's half of the court. The action of striking the shuttle is referred to as a \textit{shot}, which is the smallest unit of analysis in this study. Shots hit downward towards the opponent's court are considered offensive, while those hit upward are considered defensive. A series of shots beginning with a service and ending when the shuttle touches the court surface of either side is termed one \textit{round}.

Serving rules by the Badminton World Federation (BWF, 2023)  are stringent in order to prevent the serving team from making an offensive service. Specifically, the point at which the shuttle is hit during the serve must not be higher than 1.15 metres above the court surface. As the net height is 1.55 metres, the shuttle must fly upwards after the serve in order to pass over it. The receiving team, however, have no restrictions and may step forward and hit the shuttle near the net, giving them more flexibility and positive options on their return landing area. This is especially true in the double events, where the receiving athlete may be more willing to move forward and make an offensive return in coordination with their partner. For the third shot, it has been commonly observed that when the serving athlete hits the returned shuttle above his shoulder, it is typically an offensive move. Conversely, if he hits it below his shoulder or leaves it to his partner, the shot is not as threatening. With regard to this, we define the term \textit{intercept} as the action of the serving athlete hitting the returned shuttle above his shoulder. The success in intercepting is a key factor in terms of gaining the initiative in a round for the serving team. 

Existing literature have provided descriptive statistics of the service landing areas (Zhang, Li and Jiang 2013) and the importance assessment of the returns from a kinematic perspective (Purcell 1945). Tian (2004) argues that intercepting the return is critical for the serving team in men's doubles to gain the initiative. However, none of these studies have provided a method for predicting the return landing area or examined the relationship between the services and returns, nor have they applied statistical models to perform quantitative analyses. In this study, we aim to address these gaps in literature based on a dataset extracted from 20 men's doubles badminton matches. Specifically, we focus on the first three shots of every round and
\begin{enumerate}
    \item perform exploratory data analyses to visualize the distributions of the service and return landing area,
    \item apply the generalized logit model to find significant predictors for the return landing area,
    \item apply the logistic regression model to find where a serving athlete has a higher probability to successfully intercept the returns.
\end{enumerate}\vspace{-1pc}
Through our analysis, we are able to identify: (1) the areas where services tend to concentrate, (2) the pattern of returns, and (3) the pattern of return interceptions. The results of (1) and (3) could provide the receiving team with general information about the service landing area and where their returns are likely to be intercepted, while the results of (2) may help the serving team better respond to the returning shuttles.


The structure of this paper is as follows. In Section 2, we provide detailed information on the data, including the matches from which our data are extracted and the variables obtained in each round. The terminology used in this paper are also explained. In Section 3, we visualize the distributions of the service landing area (SLA) and return landing area (RLA) and draw conclusions about their patterns. In Section 4, we use the generalized logit model to predict RLA and identify significant predictors. In Section 5, we apply the logistic regression model to examine the areas where a serving athlete has a higher probability of intercepting the return. Finally, in Section 6, we review and validate our conclusions, discuss the limitations of our study, and suggest possible future research.
\section{Data and summary statistics}

\subsection{Data extraction and terminologies}
Our analyses are based on a dataset extracted from the videos of 20 men's doubles badminton matches held between 2012 and 2022, as listed in Table \ref{2.1-tb:match}. These matches are selected due to their high-resolution video recordings and recency. Prior to the data extraction process, we consulted 10 badminton enthusiasts and 2 professionals to determine the design of our variables. We identified 20 variables in total, but only 9 of them are found to be useful in our analyses. Detailed information on the 9 variables can be found in Table \ref{2.1-tb:variables} and further explanations on three of them whose meanings are less apparent (SLA, RLA, and 3-intercept) are provided below. 

There are two types of service in badminton: short service and long service. A short service refers to hitting the shuttle forward towards the front court of the receiving team, while a long service means hitting the shuttle upwards towards the rear court of the receiving team. The reactions and strategies in response to short and long services are quite different. As short service accounts for more than 90\% of all services, we focus solely on short services in this paper. Compared to a rough definition of the SLA by Yang (2021), we propose a more precise one for short service. We divide the receiving team's front court into three areas, namely inside, middle, and outside, based on how the receiving athlete hits the shuttle. The case when serving from the left is illustrated in Figure 1 (note that the server and the receiver shall stand within diagonally opposite service courts) and the case when serving from the right is mirrored from the left. To be more specific, at the moment of the receiver hitting the shuttle, the SLA of a round is defined as:
\begin{itemize}
    \item \textit{inside} if the hitting point is between the right shoulder of the receiving athlete and the center-line;
    \item \textit{middle} if the hitting point is between the left and right shoulders of the receiving athlete;
    \item \textit{outside} if the hitting point is between the side-line and the left shoulder of the receiving athlete.
\end{itemize}\vspace{-1pc}
\

To define the RLA, we divide the serving team's court into nine areas by referring to a common partition proposed by Dang and Pan (2021) , as illustrated in Figure 2. These areas can be intuitively grouped into three paths: the left path (\textit{No.1, No.4}, and \textit{No.7}), the center path (\textit{No.2, No.5}, and \textit{No.8}), and the right path \textit{(No.3, No.6}, and \textit{No.9}). Additionally, the front court (\textit{No.1, No.2}, and \textit{No.3}), the middle court (\textit{No.4, No.5}, and \textit{No.6}), and the rear court (\textit{No.7, No.8}, and \textit{No.9}) can be identified. These groupings will be mentioned later in this paper. It is worth noting that, the shuttle is typically hit before it really touches the court surface so that the RLA and SLA in our data are mostly determined based on the shuttle's flight path rather than its true landing area.

Note that according to our experience of doubles matches, the third shot is generally made by the serving athlete rather than his partner, as the server typically stands in a more forward position and is able to hit the shuttle at a higher flight path to make the shot more aggressive. Therefore, the variable 3-intercept is designed from the perspective of the server, i.e., rounds where the server's partner being more proper to make the third shot are excluded. For the remaining rounds, we consider a round to have ``3-intercept=Yes'' if the server successfully intercepts the return, i.e., hitting the shuttle above his shoulder, and ``3-intercept=No'' if the server hits the shuttle below his shoulder or leaves it to his partner.

In addition, among the nine useful variables, the receiver's dominant hand (RDH) must be taken into account at the outset, as the RLA pattern is completely different for left-handed and right-handed players. As most players, both professionals and amateurs, are right-handed, we mirror all data for left-handed players to focus on the model for right-handed players without loss of generality. The mirroring procedures are (with an illustration in Figure 3):

\begin{itemize}
    \item if the variable \textit{Service from} = \textit{Left}, then change it to \textit{Right}, vice versa;
    \item if the variable \textit{Foot} = \textit{Left}, then change it to \textit{Right}, vice versa;
    \item if the variable \textit{RLA} = \textit{No.1}, then change it to \textit{No.3}; if \textit{RLA} = \textit{No.4}, then change it to \textit{No.6}; if \textit{RLA} = \textit{No.7}, then change it to \textit{No.9}, vise versa.
\end{itemize}\vspace{-1pc}
\

All variables are extracted manually from each round of the matches by two badminton enthusiasts independently watching the video recordings and disagreements are resolved through discussion. An example slice of the dataset is shown in Table \ref{2.1-tb:dataset} and the detailed data processing steps are demonstrated in Figure 4. 

\begin{table} [!ht]
    \caption{Matches from which the data are extracted}
    \begin{tabular}{lllr}
        Match  & Side 1 & Side 2 & \# of rounds	\\ \midrule
        2012 Olympic Semifinal & M. Boe/C. Mogensen & Chung J.S./Lee Y.D. & 119\\
        2012 Olympic Final & M. Boe/C. Mogensen & Cai Y./Fu H.F. & 73\\
        2015 BWF World Tour Finals Semifinal & Lee Y.D./Yoo Y.S. & M. Ahsan/H. Setiawan & 120\\
        2016 Olympic Final & Zhang N./Fu H.F. & Tan W.K./Goh V.S. & 113\\
        2017 World Championship 1/4 Final & Chai B./Hong W. & M.F. Gideon/K.S. Sukamuljo & 114\\
        2018 All England Open Final & M. Boe/C. Mogensen & M.F. Gideon/K.S. Sukamuljo & 77\\
        2018 China Open Final & He J.T./Tan Q. & M.F. Gideon/K.S. Sukamuljo & 126\\
        2018 World Championship 1/4 Final & Zhang N./Liu C. & H. Endo/Y. Watanabe & 71\\
        2018 Tour Group & Liu Y.C./Li J.H. & M.F. Gideon/K.S. Sukamuljo & 81\\
        2019 All England Open 1/4 Final & Zhang N./Liu C. & M.F. Gideon/K.S. Sukamuljo & 118\\
        2019 Asia Championship Final & H. Endo/Y. Watanabe & M.F. Gideon/K.S. Sukamuljo & 63\\
        2019 China Open Final & M.F. Gideon/K.S. Sukamuljo & M. Ahsan/H. Setiawan & 113\\
        2019 Denmark Open Final & M.F. Gideon/K.S. Sukamuljo & M. Ahsan/H. Setiawan &  69\\
        2019 BWF World Tour Finals Group & M.F. Gideon/K.S. Sukamuljo & H. Endo/Y. Watanabe & 99\\
        2019 BWF World Tour Finals Semifinal & M.F. Gideon/K.S. Sukamuljo & H. Endo/Y. Watanabe & 164\\
        2019 BWF World Tour Finals Final & M. Ahsan/H. Setiawan & H. Endo/Y. Watanabe & 86\\     
        2020 All England Open Final & M.F. Gideon/K.S. Sukamuljo & H. Endo/Y. Watanabe & 112\\
        2020 Malaysia Open Final & Liu Y.C./Li J.H. & M. Ahsan/H. Setiawan & 128\\
        2022 All England Open Semifinal & M.S. Fikri/B. Maulana & M.F. Gideon/K.S. Sukamuljo & 110\\
        2022 All England Open Final & M.S. Fikri/B. Maulana & M. Ahsan/H. Setiawan & 74\\ \midrule
        Total & --- & --- & 2030 \\
    \end{tabular}
    \label{2.1-tb:match}
\end{table}

\begin{table} [!ht]
    \caption{List of variables}
    \begin{tabular}{llr}
        Variable Name & Meaning & Value\\ \midrule
        Server & The serving athlete's name of this round & ---\\
        Service from & Whether the server is serving from the left or right court & Left/Right\\
        Service landing area (SLA) & The service landing area & Inside/Middle/Outside\\
        Receiver & The receiving athlete's name of this round & ---\\
        Receiver's dominant hand (RDH) & Which hand the receiver is using to grip racket  & Left/Right\\
        Foot & With which foot the receiver step out for the first step & Left/Right\\
        Grip & How the receiver grip the racket & Forehand/Backhand\\
        Return landing area (RLA) & The return landing area & $\{$1,2,$\cdots$,9$\}$\\
        3-intercept & Whether the server intercepts the return successfully & Yes/No\\
    \end{tabular}
    \label{2.1-tb:variables}
\end{table}

\begin{table} [!ht]
    \caption{Example of the dataset}
    \begin{tabular}{llllllllr}
    Server & Service from & SLA & Receiver & RDH & Foot & Grip & RLA & 3-intercept\\\midrule
    K.S. Sukamuljo & Left & Inside & M. Boe & Left & Right & Forehand & 5 & No\\
    M.F. Gideon & Right & Middle & C. Mogensen & Right & Left & Forehand & 5 & No\\
    M.F. Gideon & Right & Middle & C. Mogensen & Right & Right & Forehand & 9 & No\\
    \end{tabular}\label{2.1-tb:dataset}
\end{table}

\subsection{Services are concentrated near the center-line}
To identify the service pattern, we summarize the distribution of SLA in Table \ref{2.2-tb:SLA}. Despite the difference between serving from the left and the right, there is a noticeable pattern in the distribution of SLA in our dataset: the majority of services (92.7\%) are concentrated either in the inside or the middle areas, which are located near the center-line as shown in Figure 1. In contrast, only a small proportion of services (7.3\%) fall into the outside area, despite the fact that it is the largest of the three areas. Similar findings have been reported by Zhang, Li, and Jiang (2013) and we will delve further into this phenomenon in Section 4.

\begin{table} [!ht]
\caption{Distribution of the service landing area}
\begin{tabular}{lccc}
& Inside & Middle & Outside \\ \midrule
Serving from the left &  571  & 304  & 6 \\
Serving from the right & 406  & 363  & 126  \\\midrule
Total & 977 (55.5\%) & 667 (37.1\%) & 132 (7.3\%) \\
\end{tabular}
\label{2.2-tb:SLA}
\end{table}

\subsection{Returns are concentrated on the crossing areas}
The distribution of RLA is summarized in Table \ref{2.3-tb:RLA} and further visualized in Figure 5 which provides a direct and clear 
understanding of the distribution of RLA in our data. We find that the majority of returns (83.6\%) fall in the crossing areas (\textit{RLA No.2, No.4, No.5, No.6} and \textit{No.8}) while a relatively small proportion of returns (16.4\%) are directed towards the corner areas (\textit{RLA No.1, No.3, No.7}, and \textit{No.9}).

\begin{table} [!ht]
\caption{Distribution of the return landing area}
\begin{tabular}{lccccccccc}
 & No.1 & No.2 & No.3 & No.4 & No.5 & No.6 & No.7 & No.8 & No.9 \\ \midrule
Serving from left & 19 & 213 & 16 & 83 & 192 & 134 & 46 & 100 & 11 \\
Serving from right & 93 & 148 & 20 & 175 & 197 & 65 & 32 & 67 & 30 \\\midrule
\multirow{2}{*}{Total} & 112 & 361 & 36 & 258 & 389 & 199 & 78 & 167 & 41 \\
& (6.8\%) & (22.0\%) & (2.2\%) & (15.7\%) & (23.7\%) & (12.1\%) & (4.8\%) & (10.2\%) & (2.5\%) \\
\end{tabular}
\label{2.3-tb:RLA}
\end{table}

\section{Prediction of the return landing area}
In this section, we would like to investigate whether some of the variables listed in Table \ref{2.1-tb:variables} can help predict the RLA in a round in men's doubles badminton matches. The RLA is a categorical response variable with more than two levels and no natural ordering, for which the generalized logit model is a common choice. The generalized logit model is known by a variety of other names, including multinomial logistic regression, multiclass logistic regression, and multinomial logit model, etc. As shown in Table \ref{2.3-tb:RLA}, the distribution of RLA may differ when serving from the left or the right. Hence, we propose to consider statistical models for these two situations separately. The case of serving from the left is discussed below and that of serving from the right is presented in Appendix A.1. The predictors considered in the model can be observed before the receiver hits the shuttle so that the results could potentially provide insights for the serving team to make informed decisions.

\subsection{Generalized logit model}
We select \textit{SLA}, \textit{Foot}, and \textit{Grip} as predictors in our model as they are observable before the receiver hits the shuttle. They are all categorical variables and we set the corresponding reference levels as \textit{SLA = Inside}, \textit{Foot = Right}, and \textit{Grip = Forehand}. Let ($i$ refers to the $i$th round of the data, $i=1, 2, \ldots, n$)
\begin{equation*}
\boldsymbol{x}_{i} = \left(  
    \begin{array}{c}
        \mbox{whether } SLA_i = Outside\\ \mbox{whether } SLA_i = Middle\\ 
        \mbox{whether } Foot_i = Left\\ \mbox{whether } Grip_i = Forehand
    \end{array}
\right), \quad \boldsymbol{Y}_i = \left(  
    \begin{array}{c}
        Y_{i1}\\ Y_{i2}\\ \vdots \\ Y_{i9}
    \end{array}
\right),
\end{equation*}
where $Y_{ij} = I(RLA_i = No.j)$ and $\sum_{j=1}^9 Y_{ij} = 1$. Let $p_{ij} = \Pr(Y_{ij}=1\mid \boldsymbol{x}_i)$, then $\sum_{j=1}^9 p_{ij}=1$. To apply the generalized logit model, it is assumed that $\boldsymbol{Y}_1, \boldsymbol{Y}_2, \ldots, \boldsymbol{Y}_n$ are independent and
\begin{equation}
    \boldsymbol{Y}_i \sim \mbox{Multinomial}
    \left(1,\hspace{0.3ex} \left(
        \begin{array}{c}
            p_{i1}\\ p_{i2}\\ \vdots \\ p_{i9}
        \end{array}
    \right)
\right).\label{gen-logit-model-ass1}
\end{equation}
The model is 
\begin{equation}
    \mbox{glogit}(p_{ij}) \triangleq \log\left(\frac{p_{ij}}{p_{i5}}\right)=\alpha_j+{\boldsymbol{x}}_{i}^\top{\boldsymbol{\beta}}_j,\quad \boldsymbol{\beta}_j=\begin{pmatrix}
        \beta_{j1} \\ \beta_{j2} \\ \beta_{j3} \\ \beta_{j4}
    \end{pmatrix}, \quad j = 1, 2, \ldots, 9.
    \label{gen-logit-model-ass2}
\end{equation}
Note that \textit{RLA = No.5} is chosen as the reference level to define the generalized logits (i.e., $\mbox{glogit}(p_{ij})$) because the returns are concentrated on the crossing areas as shown by Table \ref{2.3-tb:RLA} and Figure 5. The parameter estimates obtained by fitting the model on our data are provided in Table \ref{3.1-tb:reg-coef}. Only the parameters with p-value less than 0.2 are listed for better interpretation. The estimated probabilities for each RLA under different combinations of \textit{SLA}, \textit{Foot}, and \textit{Grip} are provided in Table \ref{3.1-tb:est-prob}.

\begin{table} [!ht]
\caption{Parameter estimates of the prediction model for RLA (when serving from the left)}
    \begin{tabular}{lcccr}
	    	Variable & RLA & Parameter & Estimate (SE) & p-value \\
	    	\midrule
    		\textit{SLA} = \textit{Outside} & \textit{No.9} & $\beta_{91}$ & 1.9393(1.0851) & 0.0739\\
    		\textit{Foot} = \textit{Left} & \textit{No.1} & $\beta_{13}$ & -1.2913 (0.5369) & 0.0162 \\
    		\textit{Foot} = \textit{Left} & \textit{No.2} & $\beta_{23}$ & -0.4150 (0.1058) & <0.0001 \\
    		\textit{Foot} = \textit{Left} & \textit{No.3} & $\beta_{33}$ & -0.8088 (0.2806) & 0.0039 \\
    		\textit{Foot} = \textit{Left} & \textit{No.6} & $\beta_{63}$ & -0.4998 (0.1188) & <0.0001 \\
    		\textit{Foot} = \textit{Left} & \textit{No.7} & $\beta_{73}$ & 0.7541 (0.2404) & 0.0017 \\
    		\textit{Foot} = \textit{Left} & \textit{No.8} & $\beta_{83}$ & -0.2347 (0.1301) & 0.0711 \\
    		\textit{Foot} = \textit{Left} & \textit{No.9} & $\beta_{93}$ & 0.5874 (0.3814) & 0.1924 \\
    		\textit{Grip} = \textit{Forehand} & \textit{No.1} & $\beta_{14}$ & -2.1315 (0.4407) & <0.0001 \\
    		\textit{Grip} = \textit{Forehand} & \textit{No.2} & $\beta_{24}$ & -0.3108 (0.2388) & 0.1930 \\
    		\textit{Grip} = \textit{Forehand} & \textit{No.4} & $\beta_{44}$ & -2.0967 (0.2628) & <0.0001 \\
    		\textit{Grip} = \textit{Forehand} & \textit{No.6} & $\beta_{64}$ & 0.8335 (0.4514) & 0.0648 \\    		
    		\textit{Grip} = \textit{Forehand} & \textit{No.7} & $\beta_{74}$ & -1.1550 (0.3311) & 0.0005 \\
    		\textit{Grip} = \textit{Forehand} & \textit{No.8} & $\beta_{84}$ & 0.8339 (0.5445) & 0.1257 \\
    	\end{tabular}\label{3.1-tb:reg-coef}
\end{table}

\begin{table} [!ht]
\caption{Estimated probability of each RLA (when serving from the left)}
\begin{tabular}{lllccccccccc}
& & & \multicolumn{9}{c}{RLA}\\
\cmidrule(r){4-12}
\textit{SLA} & \textit{Foot} & \textit{Grip} & \textit{No.1} & \textit{No.2} & \textit{No.3} & \textit{No.4} & \textit{No.5} & \textit{No.6} & \textit{No.7} & \textit{No.8} & \textit{No.9}\\
\midrule
\textit{Inside} & \textit{Right} & \textit{Backhand} & 0.17 & 0.17 & 0.00 & 0.55 & 0.05 & 0.01 & 0.04 & 0.01 & 0.00\\
& & \textit{Forehand} & 0.01 & 0.35 & 0.01 & 0.03 & 0.20 & 0.25 & 0.02 & 0.13 & 0.00\\
& \textit{Left} & \textit{Backhand} & 0.01 & 0.06 & 0.00 & 0.70 & 0.05 & 0.00 & 0.17 & 0.00 & 0.00\\
& & \textit{Forehand} & 0.00 & 0.23 & 0.00 & 0.07 & 0.31 & 0.14 & 0.11 & 0.13 & 0.01\\
\hline
\textit{Outside} & \textit{Right} & \textit{Backhand} & 0.40 & 0.19 & 0.00 & 0.00 & 0.15 & 0.12 & 0.00 & 0.00 & 0.14\\
& & \textit{Forehand} & 0.01 & 0.10 & 0.00 & 0.00 & 0.16 & 0.68 & 0.00 & 0.00 & 0.06\\
& \textit{Left} & \textit{\textit{Backhand}} & 0.04 & 0.11 & 0.00 & 0.00 & 0.19 & 0.06 & 0.00 & 0.00 & 0.60\\
& & \textit{Forehand} & 0.00 & 0.07 & 0.00 & 0.00 & 0.25 & 0.40 & 0.00 & 0.00 & 0.28\\
\hline
\textit{Middle} & \textit{Right} & \textit{Backhand} & 0.21 & 0.23 & 0.02 & 0.40 & 0.07 & 0.02 & 0.02 & 0.01 & 0.01\\
& & \textit{Forehand} & 0.01 & 0.30 & 0.10 & 0.01 & 0.18 & 0.24 & 0.01 & 0.16 & 0.01\\
& \textit{Left} & \textit{Backhand} & 0.02 & 0.11 & 0.00 & 0.63 & 0.08 & 0.01 & 0.11 & 0.01 & 0.03\\
& & \textit{Forehand} & 0.00 & 0.22 & 0.03 & 0.04 & 0.30 & 0.15 & 0.04 & 0.17 & 0.04\\
\end{tabular}\label{3.1-tb:est-prob}
\end{table}

\subsection{Result interpretations}
The first row in Table \ref{3.1-tb:reg-coef} shows $\hat{\beta}_{91}=1.9393$, suggesting that 
for \textit{SLA = Outside} the estimated ratio of the probabilities of \textit{RLA = No.9} and \textit{RLA = No.5} is $\exp (1.9393)\approx 6.95$ times that for \textit{SLA = Inside}, controlling for \textit{Foot} and \textit{Grip}. 
More specifically, the estimated probabilities of \textit{RLA = No.9} in Table \ref{3.1-tb:est-prob} is much higher under \textit{SLA = Outside} than those under \textit{SLA = Inside}, for all four combinations of \textit{Foot} and \textit{Grip}. 
We also provide Figure 6a to better illustrate this result that the receiver is more likely to return the shuttle to \textit{RLA No.9} if the service goes to the outside area. 
The flight path of a shuttle would be along the side line as shown in Figure 6a to reach \textit{RLA No.9} from \textit{SLA Outside}. 
In this case, one of the serving team players must move close to the back boundary line of the court to make the third shot, resulting in a disadvantage for the serving team. 
This is probably the reason why services rarely go to the outside area as shown in Table \ref{2.2-tb:SLA}.

Then for \textit{Foot = Left}, the estimated regression coefficients are positive for \textit{RLA No.7, No.9} ($\beta_{73}$ and $\beta_{93}$) and negative for \textit{RLA No.1 – No.3, No. 6, No.8} ($\beta_{13}$, $\beta_{23}$, $\beta_{33}$, $\beta_{63}$, and $\beta_{83}$). 
Therefore, for \textit{Foot = Left} the estimated ratio of the probabilities of \textit{RLA = No.7} (or \textit{RLA = No.9}) and \textit{RLA = No.5} is $\exp (0.7541)\approx 2.13$ (or $\exp(0.5874)\approx 1.80$) times that for \textit{Foot = Right}, controlling for \textit{SLA} and \textit{Grip}. 
The estimated probabilities of \textit{RLA = No.7} and \textit{No.9} in Table \ref{3.1-tb:est-prob} are also higher under \textit{Foot = Left} than those under \textit{Foot = Right}, for all combinations of \textit{SLA} and \textit{Grip}. 
Figure 6b displays these results with red boxes representing the two RLAs (\textit{No.7} and \textit{No.9}) with relatively higher probabilities and green boxes marking the five RLAs with relatively lower probabilities if the receiver steps out first on the left foot rather than the right foot.
Generally speaking, the receiver has a higher probability to return to the serving team’s rear court and a lower probability to return to the serving team’s front court when he steps out on the left foot first, compared to the case when he steps out on the right foot first.

Lastly, for \textit{Grip = Forehand}, the estimated regression coefficients are positive for \textit{RLA No.6, No.8} ($\hat{\beta}_{64}=0.8335, \hat{\beta}_{84}=0.8339$) and negative for \textit{RLA No.1, No.2, No.4, No.7}. Similar interpretations can be obtained as above and are displayed in Figure 6c, suggesting that the receiver has a higher probability to return to the right path of the serving team and a lower probability to return to the left path of the serving team when he uses a forehand grip, compared to the case when he uses a backhand grip.


\subsection{Suggestions for the serving team}
Summarized from the interpretations above, there are three major findings by applying the generalized logit model:
\begin{enumerate}
    \item The likelihood of a receiver returning a service along the side line increases when the service is directed towards the side line.
    \item A receiver is more likely to return a service to the \textit{rear court} of the serving team when he steps out with the left foot first than the case when he steps out with the right foot first.
    \item A receiver is more likely to return a service to the \textit{right path} of the serving team when he uses a forehand grip than the case when he uses a backhand grip.
\end{enumerate}

Based on these findings, four suggestions are offered to the serving team to facilitate instant decision-making. They are presented following the sequential order in which the three predictors (\textit{SLA}, \textit{Foot}, \textit{Grip}) are observed.

\begin{enumerate}
    \item The server should communicate with his partner about his intended service placement and ask his partner to pay particular attention to the corresponding side line if he plan to serve to the outside area.
    \item If the receiver steps out with his left foot first, the serving team should focus more on the middle and rear court, whereas if the receiver steps out with his right foot first, they should focus more on the middle and front court.
    \item If the receiver uses a forehand grip, the serving team should focus more on the center and right path, while if the receiver uses a backhand grip, they should focus more on the center and left path.
\end{enumerate}

However, in practice, excellent players are able to return services to varying areas with little difference in action. Therefore, the suggestions above are statistically valid, but not necessarily effective in every specific round.
\section{Where returns are intercepted?}
In this section, we aim to explore when a server has a higher probability of successfully intercepting the receiver's return. This might be useful for the receiving team, as it could help them take action to avoid being intercepted. To this end, we first present the interception rate of the players in our data and then proceed to model the probability of interception in a round.

\subsection{Excellent interceptors: M.S. Fikri/B. Maulana and M. Boe/C. Mogensen}
Nine out of the 25 players in our study had an interception rate greater than 30\%, as listed in Table \ref{4.1-tb:intercept-rate}. Among these nine players, M.S. Fikri and B. Maulana, as well as M. Boe and C. Mogensen, are men's double partners. These two pairs have the highest interception rate among all pairs in our data.

\begin{table} [!ht]
\caption{Players with interception rate of over 30\%}
    \begin{tabular}{lcccc}
        Player & M. Boe & B. Maulana & M.S. Fikri & Tan W.K.\\
        \midrule
        \# of interception & 27 & 19 & 19 & 10\\
        \# of service & 63 & 46 & 50 & 29 \\
        Interception rate & 42.86\% & 41.30\% & 38.00\% & 34.48\% \\
        \midrule
        \textbf{Player} & \textbf{C. Mogensen} & \textbf{Yoo Y.S.} & \textbf{K.S. Sukamuljo} & \textbf{M. Ahsan}\\
        \midrule
        \# of interception & 20 & 10 & 97 & 48\\
        \# of service & 62 & 31 & 312 & 155 \\
        Interception rate & 32.26\% & 32.25\% & 31.09\% & 30.97\% \\
        \end{tabular}
\label{4.1-tb:intercept-rate}
\end{table}

Despite the fact that all players in our data are professional players, the overall interception rate is less than 25\%, suggesting that it is challenging to intercept a return. Therefore, it is of intrinsic interest to investigate under what circumstances a return is more likely to be intercepted. 

\subsection{Logistic regression models}
Essentially, we are looking for significant variables to predict a binary response, i.e., 3-intercept. The logistic regression is widely used for its simplicity and interpretability. As above, we consider separate models for serving from the left and the right, starting with the case for serving from the left.

Let $Y_i =$ whether the return is intercepted successfully in the $i$th round (i.e., the variable 3-intercept in Table \ref{2.1-tb:variables}) and assume that
\begin{equation*}
    Y_i\stackrel{ind}{\sim} Bernoulli(p_i).
\end{equation*}

All available variables (i.e., \textit{SLA}, \textit{Foot}, \textit{Grip}, \textit{RLA}) are initially included as predictors in the logistic regression model, however, most of them are found to be insignificant. After sequential selection by the Bayesian information criterion (BIC), only \textit{RLA = Center path} and \textit{RLA = Right path} remain in the model. Hence, our final model is (with estimated regression coefficients given in Table \ref{4.2-tb:logit-reg-coef})
\begin{equation*}
     \mbox{logit}(p_i)\triangleq\log\left(\frac{p_{i}}{1-p_{i}}\right)=\beta_0+{\boldsymbol{x}}_{i}^{\top}{\boldsymbol{\beta}},\quad\mbox{where }{\boldsymbol{x}}_{i}=
    \left(  \begin{array}{c}
        \mbox{whether } RLA_i = Center \mbox{ }path\\ \mbox{whether } RLA_i = Right \mbox{ }path
        \end{array}
\right),\;\boldsymbol{\beta}=\begin{pmatrix}
    \beta_1 \\ \beta_2
\end{pmatrix}.
\end{equation*}

\begin{table} [!ht]
\caption{Parameter estimates of the fitted logistic regression model (when serving from the left)}
    \begin{tabular}{lccr}
	    	Variable & Parameter & Estimate (SE) & p-value \\
	    	\midrule
    		\textit{RLA = Center path} & $\beta_1$ & 0.4036 (0.1083) & 0.0002\\
    		\textit{RLA = Right path} & $\beta_2$ & -0.3162 (0.1485) & 0.0333\\
    	\end{tabular}
\label{4.2-tb:logit-reg-coef}
\end{table}

\subsection{Result interpretations}
The estimated regression coefficient is positive for \textit{RLA = Center path} ($\hat{\beta}_1=0.4036$) and negative for \textit{RLA = Right path} ($\hat{\beta}_2=-0.3162$), suggesting that the estimated odds of being intercepted for a return in the center path (or the right path) is $\exp(0.4036)\approx 1.5$ (or $\exp(-0.3162)\approx 0.73$) times that for a return in the left path. In other words, when serving from the left, a return is more likely to be intercepted in the center path and less likely to be intercepted in the right path, as demonstrated in Figure 6(a). In Figure 6, the baseline area is marked with a blue box and the areas within which a return has higher and lower probability of being intercepted are marked with a red and a green box, respectively. 

We provide a possible explanation for this result. The badminton men’s double is quite fast paced. After serving from the left, the return shuttle typically comes back very quickly, leaving little time for the server to move. This may be the reason why the server has a higher probability to intercept the return in the center path and left path as he just stayed there to serve.

Logistic regression models are also applied to the case when serving from the right (see Appendix A.2 for details). The results are different from those when serving from the left, as illustrated by Figure 6(b). 

\subsection{Suggestions for the receiving team}
The analysis of the two logistic regression models has led to the following three main conclusions:
\begin{enumerate}
    \item When serving from the left, servers tend to concentrate their efforts on the left and center path in an attempt to intercept the return.
    \item When serving from the right, they tend to focus more on the right path for interception.
    \item Returns towards the rear court are left to server's partner to deal with.
\end{enumerate}
As previously mentioned, the server may be more likely to miss the return when serving to the outside from the left side, due to the discrepancy between the expected return path and the server's focus on the left side, which may explain why this type of serve is not commonly used. A similar situation arises when serving from the right side. Based on the conclusions drawn from the logistic regression models, the receiving team may want to consider the following strategies:
\begin{enumerate}
    \item When their opponent is serving from the left, the receiving team could aim for \textit{RLA No.6}, as depicted in Figure 7 (a).
    \item When their opponent is serving from the right, the receiving team could aim for \textit{RLA No.4}, as depicted in Figure 7 (b).
\end{enumerate}

In fact, a review of Table \ref{2.3-tb:RLA} reveals that \textit{RLA No.4} accounts for 22.5\% of all returns when serving from the left, which is consistent with our recommendation.

\section{Conclusions, validation, limitations and further research}

\subsection{Conclusions}
To summarize our findings for readers who may not be familiar with the terminology and strategies used in badminton, we can say that serving athletes typically serve to the middle and inside areas of the receiving team's court at the start of a rally, with only a small number of serves going to the outside. On the other hand, receiving athletes often return the serve to the crossing area of the serving team's court, with only a small percentage of returns going to the corners. And after introducing statistical models, three main factors that can predict the return path of the receiving team are identified:
\begin{enumerate}
    \item If the serve is directed towards the sideline, the receiver is more likely to return the serve along the sideline.
    \item If the receiver steps out with his left foot first, he is more likely to return the service to the rear court of the serving team, while if he steps out with his right foot first, he is more likely to return to the front court.
    \item If the receiver uses a forehand grip, he is more likely to return the service to the right path, while if he uses a backhand grip, he is more likely to return to the left path.
\end{enumerate}

To intercept the opponent's return for an offensive shot, servers typically try to anticipate and focus on one direction in advance:
\begin{enumerate}
    \item When serving from the left, servers tend to intercept the return in the left or center path.
    \item When serving from the right, servers tend to intercept the return in the right or center path.
    \item If the return is directed towards the rear court, servers are less likely to intercept and leave it to their partners.
\end{enumerate}

\subsection{Validation}
To validate our findings, we have recorded a match from the Malaysia Masters 1/8 Final in 2022, featuring the teams of Wang C./Liang W.K. (China) and Takuro Hoki/Yugo Kobayashi. These two pairs are not included in our original dataset. After excluding rounds that began with faulty, error, or long services, we have a total of 89 rounds for analysis.

Of the 89 short services in this match, only one went to the outside area (first game, 20:14) and was returned along the sideline, which supports our conclusion. However, because the number of services to the outside area was so small, additional data would be needed to confirm this finding.

We will now examine the second predictor, \textit{Foot}, which classifies whether the return goes to the rear court or the front court. Specifically, we will test the following logic: if the receiver steps out with their left foot first, the return is more likely to go to the rear court of the serving team; if the receiver steps out with their right foot first, the return is more likely to go to the front court of the serving team. A contingency table based on the validation data is presented in Table \ref{5.2-tb:foot-valid}.

\begin{table} [!ht]
\caption{Contingency table for the variable \textit{Foot}}
    \begin{tabular}{lccc}
	    	 & Front court & Middle court & Rear court \\
	    	\midrule
    		 Left foot & 2 & 13 & 12\\
    		 Right foot & 22 & 24 & 15\\
    		 Total & 24 & 37 & 27\\
    	\end{tabular}
\label{5.2-tb:foot-valid}
\end{table}
If we consider only the cases where the receiver steps out with their right foot and the return goes to the front court, or the receiver steps out with their left foot and the return goes to the rear court, as correct classifications, our accuracy would be approximately 38.6\% based on the validation data. However, if we also consider the middle court as a potential return destination for either foot, the accuracy increases to approximately 80.7\%. This suggests that focusing on just one part of the court can be risky, and considering multiple potential return areas is more reasonable. Therefore, we conclude that the \textit{Foot} variable is a useful indicator for predicting whether the return will go to the front or rear court.

The final predictor to be examined is \textit{Grip}, which indicates whether the return will go to the serving team's right or left path. The classification logic is as follows: if the receiver uses a forehand grip, the return is more likely to go to the right path, and if the receiver uses a backhand grip, the return is more likely to go to the left path. The contingency table for this predictor is shown in Table \ref{5.2-tb:grip-valid}.

\begin{table} [!ht]
\caption{Contingency table for grip classifier}
    \begin{tabular}{lccc}
	    	 & Left path & Center path & Right path \\
	    	\midrule
    		 Forehand grip & 5 & 21 & 15\\
    		 Backhand grip & 24 & 19 & 4\\
    		 Total & 29 & 40 & 19\\
    	\end{tabular}
\label{5.2-tb:grip-valid}
\end{table}

Similarly to the \textit{Foot} predictor, it is not advisable to focus on only one return path based on the receiver's grip. Instead, we suggest focusing on the center path and using the grip as an additional indicator to predict the direction. Using this logic, the accuracy would be approximately 89.8\% based on the validation data. Therefore, we conclude that the \textit{Grip} variable is also a useful predictor.

\subsection{Limitation and further research}
There are a few limitations to our study that should be acknowledged. One significant limitation is the relatively small sample size of 2030 rounds, considering that the data was collected from 25 athletes. This may not be sufficient for a general analysis. Another limitation is the method used to record the landing areas, which is based on the shuttle's flight path rather than the actual landing spot. This may introduce some bias into the data, as athletes may attempt to prevent the shuttle from hitting the ground.

Throughout this study, we have observed that different athletes tend to have slightly different preferences for which foot to step with first and which grip to use. Therefore, we believe that our method could potentially be more accurate if it is applied specifically to a single athlete or doubles pair, rather than being generalized to all athletes.

To date, there have been relatively few research studies focused on badminton compared to sports such as basketball or soccer. As a result, when conducting our research, we found limited relevant studies to reference and had to create many new terms and variables through a focus group meeting. We hope that more professionals and scholars will become interested in this field, potentially leading to the development of more effective methods for describing and recording situations on the badminton court, as well as more sophisticated models for capturing the key features of athletes' actions.
\begin{appendix}
\section{Appendix}
\subsection{Generalized logit models for RLA prediction when serving from the right}
Assumptions are the same as Equation (\ref{gen-logit-model-ass1}) and (\ref{gen-logit-model-ass2}) in the generalized logit model in Section 3.1. Table \ref{app-tb1:reg-coef} shows the parameter estimates obtained by fitting the model on the data extracted. Note that only the parameters with p-value less than 0.2 are listed. The estimated probabilities for each RLA under different combinations of SLA, Foot and Grip are provided in Table \ref{app-tb2:est-prob}.
\begin{table} [!ht]
\caption{Parameter estimates of the prediction model for RLA (when serving from the right)}
    \begin{tabular}{lcrrr}
	    	Variable & RLA & Parameter & Estimate(SE) & p-value \\
	    	\midrule
    		\textit{SLA} = \textit{Outside} & \textit{No.1} & $\beta_{11}$ & 2.1379 (0.3712) & <0.0001\\
    		\textit{SLA} = \textit{Outside} & \textit{No.4} & $\beta_{41}$ & 1.2577 (0.3417) & 0.0002\\
    		\textit{SLA} = \textit{Outside} & \textit{No.6} & $\beta_{61}$ & -0.7191 (0.3269) & 0.0278\\
    		\textit{SLA} = \textit{Outside} & \textit{No.8} & $\beta_{81}$ & 0.7007 (0.3207) & 0.0289\\
    		\textit{SLA} = \textit{Middle} & \textit{No.1} & $\beta_{12}$ & 0.3650 (0.2330) & 0.1173\\
    		\textit{SLA} = \textit{Middle} & \textit{No.6} & $\beta_{62}$ & 0.7481 (0.2200) & 0.0007\\
    		\textit{Foot} = \textit{Left} & \textit{No.1} &$\beta_{13}$ &  -0.3512 (0.1623) & 0.0305 \\
    		\textit{Foot} = \textit{Left} & \textit{No.4} & $\beta_{43}$ & 0.1677 (0.1152) & 0.1454 \\
    		\textit{Foot} = \textit{Left} & \textit{No.7} & $\beta_{73}$ & 0.7034 (0.2223) & 0.0016 \\
    		\textit{Foot} = \textit{Left} & \textit{No.8} & $\beta_{83}$ & 0.2297 (0.1510) & 0.1281 \\
    		\textit{Foot} = \textit{Left} & \textit{No.9} & $\beta_{93}$ & 0.9031 (0.2536) & 0.0004 \\
    		\textit{Grip} = \textit{Forehand} & \textit{No.1} & $\beta_{14}$ &  -1.5421 (0.2547) & <0.0001 \\
    		\textit{Grip} = \textit{Forehand} & \textit{No.2} & $\beta_{24}$ & -0.3116 (0.1476) & 0.0347 \\
    		\textit{Grip} = \textit{Forehand} & \textit{No.3} & $\beta_{34}$ & 0.5340 (0.2817) & 0.0580 \\
    		\textit{Grip} = \textit{Forehand} & \textit{No.4} & $\beta_{44}$ & -1.6042 (0.2547) & <0.0001 \\    	
    		\textit{Grip} = \textit{Forehand} & \textit{No.6} & $\beta_{64}$ & 0.9594 (0.1985) & <0.0001 \\
    		\textit{Grip} = \textit{Forehand} & \textit{No.7} & $\beta_{74}$ & -0.8388 (0.3208) & 0.0089 \\
    		\textit{Grip} = \textit{Forehand} & \textit{No.8} & $\beta_{84}$ & -0.3905 (0.2003) & 0.0512 \\
    		\textit{Grip} = \textit{Forehand} & \textit{No.9} & $\beta_{94}$ & 0.3056 (0.2332) & 0.1900 \\
    	\end{tabular}\label{app-tb1:reg-coef}
\label{tab:Table11}
\end{table}

\begin{table} [!ht]
\caption{Estimated probability of each RLA (when serving from the right)}
\begin{tabular}{lllccccccccc}
& & & \multicolumn{9}{c}{RLA}\\
\cmidrule(r){4-12}
SLA & Foot & Grip & No.1 & No.2 & No.3 & No.4 & No.5 & No.6 & No.7 & No.8 & No.9\\
\midrule
Inside & Right & Backhand & 0.04 & 0.31 & 0.02 & 0.20 & 0.29 & 0.02 & 0.03 & 0.08 & 0.01\\
& & Forehand & 0.00 & 0.23 & 0.08 & 0.01 & 0.41 & 0.18 & 0.01 & 0.05 & 0.04\\
& Left & Backhand & 0.02 & 0.22 & 0.01 & 0.23 & 0.24 & 0.02 & 0.09 & 0.10 & 0.07\\
& & Forehand & 0.00 & 0.16 & 0.05 & 0.01 & 0.32 & 0.19 & 0.02 & 0.06 & 0.18\\
\hline
Outside & Right & Backhand & 0.55 & 0.04 & 0.00 & 0.33 & 0.04 & 0.00 & 0.00 & 0.03 & 0.00\\
& & Forehand & 0.20 & 0.15 & 0.03 & 0.11 & 0.30 & 0.07 & 0.01 & 0.13 & 0.02\\
& Left & Backhand & 0.30 & 0.04 & 0.00 & 0.52 & 0.04 & 0.00 & 0.02 & 0.06 & 0.01\\
& & Forehand & 0.09 & 0.12 & 0.02 & 0.13 & 0.26 & 0.08 & 0.02 & 0.17 & 0.10\\
\hline
Middle & Right & Backhand & 0.32 & 0.11 & 0.01 & 0.33 & 0.13 & 0.02 & 0.02 & 0.05 & 0.00\\
& & Forehand & 0.04 & 0.15 & 0.07 & 0.03 & 0.32 & 0.31 & 0.01 & 0.06 & 0.01\\
& Left & Backhand & 0.15 & 0.09 & 0.01 & 0.44 & 0.12 & 0.02 & 0.08 & 0.07 & 0.02\\
& & Forehand & 0.01 & 0.11 & 0.05 & 0.04 & 0.27 & 0.35 & 0.03 & 0.07 & 0.07\\
\end{tabular}\label{app-tb2:est-prob}
\end{table}

The first variable in Table \ref{app-tb1:reg-coef} is \textit{SLA = Outside} for \textit{RLA = No.1, No.4, No.6, No.8}. Since $\hat{\beta}_{11}=2.1379$ and $\hat{\beta}_{41}=1.2577$, for \textit{SLA = Outside} the estimated odds of \textit{RLA = No.1} (or \textit{RLA = No.4}) rather than \textit{RLA = No.5} is $\exp(2.1379)\approx 8.48$ (or $\exp(1.2577)\approx 3.52$). And Table \ref{app-tb2:est-prob} further confirms this result, where the estimated probabilities for \textit{SLA = No.1, No.4} under \textit{SLA = Outside} are generally higher than those when \textit{SLA = Inside}. Figure 8(a) is provided to better illustrate the implication  that returns are more likely to be along the sideline when the service is directed to the outside area, consistent with our earlier conclusion for services from the left.

Moreover, the model shows that the variable \textit{Foot = Left} is a strong predictor for returns to the rear court of the serving team, with estimated coefficients that are consistently positive for the whole rear court (\textit{RLA No.7, No.8, No.9}). So, we may conclude the result that the returns are more likely to be directed to the rear court of the serving team as depicted in Figure 8(b). However, the estimated probabilities in Table \ref{app-tb2:est-prob} for the rear court do not seem significantly higher when the receiver steps out on the left foot first comparing to the case when he steps out on the right foot first.

To conclude, the variable \textit{Grip = Forehand} is a good predictor of the path that the return will take, with the probability of a return to the right path of the serving team being higher when the receiver uses a forehand grip as shown in Table 8(c). This is supported by the estimated coefficients for the right path being positive and those for the left path being negative. The estimated probability differences between the left path and the right path when the receiver uses a forehand grip compared to a backhand grip are also noticeable, as can be seen in Table \ref{app-tb2:est-prob}.

\subsection{Logistic regression models for return interceptions when serving from the right}
Assume (subscript i refers to the $i$-th row of the data)
\begin{equation*}
    Y_i\stackrel{ind}{\sim} Bernoulli(p_i),
\end{equation*}
and
\begin{equation*}
     \mbox{logit}(p_i)\triangleq\log\left(\frac{p_{i}}{1-p_{i}}\right)=\beta_0+{\boldsymbol{x}}_{i}^{\top}{\boldsymbol{\beta}},\quad\mbox{where }{\boldsymbol{x}}_{i}=
    \left(  \begin{array}{c}
        \mbox{whether \textit{RLA} = \textit{No.1}}\\ \mbox{whether \textit{RLA} = \textit{No.2}}\\
        \vdots\\
        \mbox{whether \textit{RLA} = \textit{No.9}}\\
        \end{array}
\right),\;\boldsymbol{\beta}=\begin{pmatrix}
    \beta_1 \\ \beta_2 \\ \vdots \\ \beta_9
\end{pmatrix}.
\end{equation*}

\begin{table} [!ht]
\caption{Significant variables in the intercepting model (right)}
    \begin{tabular}{lccr}
	    	Variable & Parameter & Estimate(SE) & p-value \\
	    	\midrule
    		 \textit{RLA = No.1} & $\beta_1$ & 1.0218 (0.2646) & 0.0001\\
    		 \textit{RLA = No.2} & $\beta_2$ & 1.4526 (0.2366) & <0.0001\\
    		 \textit{RLA = No.3} & $\beta_3$ & 1.4236 (0.4931) & 0.0039\\
    		 \textit{RLA = No.6} & $\beta_6$ & 1.0669 (0.3052) & 0.0005\\
    		 \textit{RLA = No.7} & $\beta_7$ & -2.0422 (0.9167) & 0.0259\\
    		 \textit{RLA = No.8} & $\beta_8$ & -1.4981 (0.4912) & 0.0023\\
    		 \textit{RLA = No.9} & $\beta_9$ & -1.7545 (0.9218) & 0.0570\\
    	\end{tabular}
\label{app-tb3:logit-reg-coef}
\end{table}

The estimated regression coefficients are listed in Table \ref{app-tb3:logit-reg-coef}. Based on this result, it appears that the server is more likely to choose to intercept returns to the front court, as indicated by the positive estimated coefficients for the right path (\textit{RLA = No.1, No.2, No.3} with values of $\hat{\beta}_1=1.02, \hat{\beta}_2=1.45, \hat{\beta}_3=1.42$) and the negative estimated coefficients for the left path (\textit{RLA = No.7, No.8, No.9} with the values of $\hat{\beta}_7=-2.04, \hat{\beta}_8=-1.50, \hat{\beta}_9=-1.75$). The estimated odds of an successful interception for a return to \textit{RLA No.1} (or \textit{RLA No.7}) is $\exp(1.02)\approx 2.77$ (or $\exp(-2.04)\approx 0.14$) times that for a return to \textit{RLA No.5} and so on so forth. For better illustration, the plot is given in Section 4.3. 
\end{appendix}



\end{document}